\documentclass[12pt]{article}
\usepackage{epsfig}
\usepackage{psfrag}
\usepackage{amsmath}
\usepackage{lscape}
\usepackage{chemarr}
\usepackage{multirow}
\usepackage{graphics}
\usepackage[super, comma, sort&compress]{natbib}
\newcommand{\bfig}{\begin{figure}[h] \begin{center}}
\newcommand{\efig}{\end{center} \end{figure}}

\begin{document}

\begin{center}
 Heterogeneous path ensembles for conformational transitions in 
semi--atomistic models of adenylate kinase 
\end{center}

\begin{center}
Divesh Bhatt, Daniel M. Zuckerman\footnote{email: ddmmzz@pitt.edu}
\end{center}
\begin{center}
Department of Computational Biology,
University of Pittsburgh
\end{center}

\centerline{\Large \bf Abstract}

 We performed ``weighted ensemble'' path--sampling simulations of adenylate
kinase, using several semi--atomistic protein models. Our study investigated
both the biophysics of conformational transitions as well as the possibility
of increasing model accuracy without sacrificing good sampling. Biophysically,
the path ensembles show significant heterogeneity and the explicit possibility
of two principle pathways in the Open$\leftrightarrow$Closed transition.
We recently showed, under certain conditions, a 
``symmetry of hereteogeneity'' is
expected between the forward and the reverse transitions: the fraction of
transitions taking a specific pathway/channel will be the same in both
the directions. Our path ensembles are analyzed in the light of the symmetry
relation and its conditions. In the realm of modeling, we employed an
all--atom backbone with various levels of residue interactions. Because
reasonable path sampling required only a few weeks of single--processor
computing time with these models, the addition of further chemical detail
should be feasible.

\section{Introduction}
\label{sec:intro}

 Fluctuations and conformational changes are of extreme importance 
in biomolecules.\cite{biochem}
For example, most enzymes show 
distinctly different conformations in the apo and the holo forms.\cite{enz}
Conformational transitions are also typical in non--enzymatic
binding proteins,\cite{biochem} 
and of course are intrinsic to motor proteins.

 The fundamental biophysics of conformational transitions
in biomolecules is contained in the ensemble of paths 
-- {\it i.e.}, trajectories in configurational space -- defining
the transition. Such path ensembles contain the information
about the relevant ``mechanisms'' for transitions, including  possible
intermediates. In addition, the transition rates can only
be calculated accurately from a path ensemble, which implicitly accounts
for all barriers and recrossings.\cite{tps2}
From a computational point of view, such path ensembles are
difficult to obtain due to rugged energy landscapes and the 
timescales involved.\cite{tps2,tps4,tis1,tis2,mile1,mile2,mile3,ffs1,ffs2,ffs5}
Multiple local minima and/or channels dramatically increase the computational
effort required. To put the difficulty of path sampling in perspective,
note that equilibrium sampling of fully atomistic
models of large biomolecules is not typically feasible.\cite{ed2} Thus, path
sampling using detailed atomistic models for all, but the smallest
systems, is impractical -- even with potentially efficient methods
developed specifically for path sampling.
A number of groups have reported atomistic 
path sampling studies for small systems.\cite{trp,hb,schlick,dinner}

 Less computationally expensive approaches to determining atomistic
paths are available, including targeted and steered molecular 
dynamics,\cite{tmd1,tmd2,tmd3}
``nudged elastic band'',\cite{neb3,neb4,neb5,neb6} 
and related approaches.\cite{action1,action2,action3,action4}
However, all these
methods yield only a single path or a handful, and not the ensemble
required for a correct thermal/statistical description.
Specifically, fluctuations in pathways,
the possibility of multiple pathways (path heterogeneity), and
possible recrossings typically are not accounted for in these approaches. 

 Coarse--grained (CG) models, on the other hand, permit an alternative
strategy for statistical path sampling.\cite{cgpath2,cgpath1,dmz_dgo}
Although CG models omit
chemical detail, they can be sampled
significantly faster than fully atomistic models, and, thus, such
models are quite attractive for path sampling studies. For example,
Zhang {\it et al.}\cite{bin} showed that a simple alpha--carbon
model of calmodulin can be fully path sampled using the weighted ensemble
path sampling method.\cite{huber}
Because full path sampling in this model required only a few weeks of
single--processor computing, it is evident that better models and/or
larger systems could be studied. Network models have also been used to study
conformational transitions.\cite{adk_cg1,adk_cg3}

 In this manuscript, we report path sampling studies of adenylate kinase
which represent improvements over previous work\cite{bin} in
several ways. (i) At 214 residues, adenylate kinase is triple the
triple the size of the calmodulin domain previously path sampled. (ii) Our
models now include significant atomic details, as explained below.
(iii) We examine a series of models to test the sensitivity of the
path ensemble to the chosen interactions and parameters. (iv) We
investigate symmetry, based on our recent formal derivation,\cite{symm}
between forward and reverse transitions.

Adenylate kinase (Adk) is an enzyme that
catalyzes phosphate transfer between AMP and ATP via
\begin{equation}
\mathrm{AMP}+\mathrm{ATP}\xrightleftharpoons{\mathrm{Mg}^{2+}}
 2\mathrm{ADP},
\label{e1}
\end{equation}
and thus helps to regulate the relative amounts of 
cellular energetic units.\cite{adk_md1,eva1,adk_chu}
The crystal structure of Adk for {\it E. coli} is available
in several conformations. Its native apo form 
(Protein Databank code 4AKE\cite{adk_4ake})
is shown in Figure~\ref{fig:crystal} (a). In the figure,
the blue segments represent the core (CORE), the yellow segment
represents the AMP binding domain (BD), and the green segment represents
the flexible lid (LID). Upon ligand binding, the enzyme closes over the 
ligands.  The crystal structure (1AKE)\cite{adk_1ake}
of the holo form of the enzyme obtained in complex with
a ligand that mimics both AMP and ATP is shown
in Figure~\ref{fig:crystal} (b).
Clearly, in the apo form, the enzyme shows an Open structure (that we
denote as O in this manuscript), and in the holo form, it is Closed
(denoted by C throughout).

 Adk has been studied previously via computational methods
using both coarse--grained models and fully atomistic simulations.
Coarse--grained models used to study transition pathways
for Adk have, primarily, utilized network 
models.\cite{adk_cg1,adk_cg2,adk_cg3,adk_cg4}
In these methods,
the fluctuations in proteins are represented by
harmonic potentials, and the deformations due to these
fluctuations are used to estimate the free energy in the
basins (end states and/or multiple basins). Subsequently,
a minimum energy path is calculated to characterize the
transition.

A few groups have also studied conformational fluctuations
in Adk using atomistic models.
In an interesting amalgamation of coarse and atomistic
models, Arora and Brooks\cite{brooks_adk}
performed atomistic (with implicit
solvent) umbrella sampling molecular dynamics (MD) simulations
along an initial minimum energy path suggested by a network model.
Kubitzki and de Groot\cite{adk_rex} performed replica exchange MD 
for atomistic Adk to increase conformational sampling of
adenylate kinase -- and observed both O and C conformers;
however, a true path ensemble is not obtained from replica exchange.
In other work, fully atomistic MD on the two end structures has
been performed to observe fluctuations in the two 
ensembles\cite{adk_md1,adk_md2} but
direct conformational transitions were not observed.

 In the present study, we use semi--atomistic models to improve
chemical accuracy compared to typical coarse--grained models 
while still performing high quality path sampling.
In our models, the backbone is fully atomistic  to provide
chemically realistic geometry. Inter--residue
interactions are modeled at a coarse--grained level via the commonly used
double--G$\bar{\mathrm{o}}$ potentials\cite{dmz_dgo,hummer_dgo,adk_cg4,onu_dgo}
that (meta)stabilize two crystal structures. Additionally, one
of the models uses residue--specific interactions to probe the effect
of such interactions.  We use a library--based
Monte Carlo (LBMC) scheme to perform sampling.\cite{lbmc}. 
LBMC was previously developed
in our group and shown to facilitate the use of semi--atomistic
models of the type used here.\cite{lbmc}

 Transitions between the Open and the Closed states (both directions) 
are studied with the weighted--ensemble (WE) path--sampling method\cite{huber}
that has been previously been studied to study
folding of proteins,\cite{we_fold1} protein dimerization,\cite{we_ass}
and conformational transitions in an alpha--carbon
model of calmodulin.\cite{bin}
WE was shown to promote efficient path sampling of conformational
transitions in purely alpha--carbon model of calmodulin.\cite{bin}
Additionally, WE is statistically exact: it preserves natural system 
dynamics, resulting in an unbiased path ensemble.\cite{bin2}

 Biophysically, we focus on heterogeneity of the path ensemble (multiple
pathways) and the forward--reverse ``symmetry'' of the ensemble. It is
possible that evolution has favored the fine--tuned precision of a single
pathway in some systems, but the ``robustness'' of alternative pathways in 
other cases. Although our semi--atomistic models preclude biochemically
precise conclusions, our path sampling means that we can provide a complete
description in a model system. Further, good path sampling enables us to
investigate, perhaps for the first time, the issue of symmetry between forward
and reverse transitions -- which has implications for studies of protein
unfolding.\cite{unfold3,unfold1,unfold2}

The goal of this work, in summary, is to
probe the biophysics of transitions with the most detailed
models that allow for generating an ensemble of pathways.
The manuscript is organized as follows. First, In Section~\ref{sec:model} 
we discuss the models we use to depict the protein. 
Section~\ref{sec:method} then
describes the method to generate the ensemble of pathways.
In Section~\ref{sec:result},  we present results for transitions 
in both the directions
for all the three models we used. We discuss the results,
efficiency, and future models in Section~\ref{disc},
with conclusions given in Section~\ref{cnc}.

\section{Semi--atomistic models}
\label{sec:model}

 We use three semi--atomistic models, expanding on our previous 
work.\cite{lbmc}
In all the models, the backbone is represented in full atomistic
detail, using the three residues
alanine, glycine, and proline.\cite{lbmc}
All intraresidue interactions are included explicitly, using the OPLSAA
all--atom force field. 
Both the intra--residue interaction energies and the configurations are stored 
in libraries as described previously.\cite{lbmc}
In brief, we note that libraries of the three types
of residues are pre--generated according to the Boltzmann
distribution at 300 K, and alanine is used to represent the backbone of all 
residues besides glycine and proline (a simplification
motivated by the similarity of Ramachandran maps for the 
residues).\cite{rotamer}
Ligands are not modeled explicitly in this path sampling study.

The differences in the three models lie
in the treatment of inter--residue interactions: two of the models
use only double--G$\bar{\mathrm{o}}$ interactions at backbone alpha carbons, 
whereas one model uses both double--G$\bar{\mathrm{o}}$ and residue--specific
interactions. Complete information is given below.

All three semi--atomistic models employ double--G$\bar{\mathrm{o}}$ interactions.
Following Ref~\cite{dmz_dgo}, for each of the two crystal structures,
residues pairs with alpha carbons less than 8 $\mathrm{\AA}$ apart are
considered native contacts. In the G$\bar{\mathrm{o}}$ energy of an arbitrary configuration,
every native contact from the Open form is assigned an energy of $-\epsilon$,
whereas those exclusively found in the Closed form are scaled to be
$-e_{\mathrm{scale}}\epsilon$.
G$\bar{\mathrm{o}}$ interactions do not distinguish between
different types of residues except in terms of size.
This double--G$\bar{\mathrm{o}}$ potential between two residues $i$ and $j$ with alpha carbon
distance $r_{ij}$ is given by
\begin{equation}
u^{\mathrm{G\bar{o}}}(r_{ij})=
 \begin{cases} \infty, & r_{ij}<r_{ij}^{\mathrm{X}}(1-\delta ) \\
               -\epsilon_{\mathrm{X}}, & r_{ij}^{\mathrm{X}}(1-\delta )
                  \le r_{ij}< r_{ij}^{\mathrm{X}}(1+\delta )\\
               0.3\epsilon, & r_{ij}^{\mathrm{X}}(1+\delta )
                  \le r_{ij}< r_{ij}^{\mathrm{Y}}(1+\delta )\\
               -\epsilon_{\mathrm{Y}}, & (\ast ) r_{ij}^{\mathrm{Y}}(1-\delta )
                  \le r_{ij}< r_{ij}^{\mathrm{Y}}(1+\delta )\\
               0, & r_{ij}\ge r_{ij}^{\mathrm{Y}}(1+\delta )
 \end{cases}
\label{e3}
\end{equation}
where $r_{ij}^{\mathrm{X}}$ and $r_{ij}^{\mathrm{Y}}$ are the native distances
in the two crystal structure ordered such that
$r_{ij}^{\mathrm{X}}<r_{ij}^{\mathrm{Y}}$ (X and Y equate to Open or Closed),
and $\delta$ is a well--width parameter chosen to be 0.05. 
If X equals Open and Y equals Closed, $\epsilon_{\mathrm{X}}=\epsilon$
and $\epsilon_{\mathrm{Y}}=e_{\mathrm{scale}}\epsilon$, and vice--versa.
In the case of overlapping square wells,
$r_{ij}^{\mathrm{X}}(1+\delta )>r_{ij}^{\mathrm{Y}}(1-\delta )$, the
$0.3\epsilon$ barrier in the middle does not exist and the lower limit of
the inequality marked with $(\ast )$ is replaced by 
$r_{ij}^{\mathrm{X}}(1+\delta )$.

 The total G$\bar{\mathrm{o}}$ interactions are therefore
\begin{equation}
U_{{\mathrm{G}}\bar{\mathrm{o}}} (e_{\mathrm{scale}})= 
\sum_{i<j} u^{{\mathrm{G}}\bar{\mathrm{o}}}(r_{ij}).
\label{e4}
\end{equation}
For Model 1, we have 
\begin{equation}
U_1=U_{\mathrm{G\bar{o}}} (e_{\mathrm{scale}}=1).
\label{e5}
\end{equation}

 The motivation behind using such a double G$\bar{\mathrm{o}}$ potential is
that the two ``end'' crystal structures are presumed stable 
-- and the double--G$\bar{\mathrm{o}}$ protocol
guarantees that bistability. Such double G$\bar{\mathrm{o}}$ interactions have been used
to probe the biophysics of several 
systems.\cite{dmz_dgo,hummer_dgo,adk_cg4,onu_dgo}

\subsection{Model 1: Pure double G$\bar{\mathrm{o}}$ with energy symmetry}

Previous path sampling studies of proteins were limited to smaller systems
and/or simpler models.
We, therefore, first study whether the simplest model within the 
semi--atomistic framework can be fully path sampled. Our Model 1 
omits most chemical details and uses only symmetric 
double--G$\bar{\mathrm{o}}$ interactions as given in eq~\ref{e4}.
That is, native contacts in the Open and the Closed structure are
treated identically ($e_{\mathrm{scale}}=1$).

Because Model 1 is a pure G$\bar{\mathrm{o}}$ model,
the temperature is specified in units of the well depth of
G$\bar{\mathrm{o}}$ interactions, $\epsilon$.
We choose the temperature as the highest
at which the two experimental crystal structures are stable.
We therefore performed a series of Monte Carlo simulations, as
described below, at various temperatures.
At $T=0.75\epsilon /k_{\mathrm{B}}$,
both structures melted, but both remained (meta)stable at 
$T=0.7\epsilon /k_{\mathrm{B}}$.
Thus, for Model 1, all subsequent equilibrium and path sampling
simulations were performed at $T=0.7\epsilon /k_{\mathrm{B}}$.

\subsection{Model 2: Double G$\bar{\mathrm{o}}$ with 
residue--specific interactions}

 Our second model adds chemical detail, both to improve upon the
simplicity of Model 1, and to provide a way to check the sensitivity
of our results to modeling choices. Model 2 includes atomistic backbone
hydrogen bonding, Ramachandran propensities, and residue--specific
contact interactions, as detailed below. Because these 
interactions are implemented as
short ranged, Model 2 is only about 30$\%$ slower than Model 1, based
on wall--clock time per MC step. G$\bar{\mathrm{o}}$ interactions are again symmetric,
with $e_{\mathrm{scale}}=1$.
 
Our semi--atomistic LBMC platform makes the inclusion of additional 
interactions straightforward. Since the backbone is modeled atomistically,
backbone--backbone hydrogen bonding is easily incorporated,
as described below.
However, due to the absence of explicit side chains in 
the present implementation,
residue--specific chemical interactions can only be
incorporated at a coarse--grained level. We use residue--specific
contact interactions based on the work of Miyazawa and 
Jernigan (MJ),\cite{mj1,mj2}
as discussed below.
Specifically, we use the potential energy
\begin{equation}
U_2=U_{\mathrm{G\bar{o}}} (e_{\mathrm{scale}}=1)+U_{\mathrm{HB}}
 +U_{\mathrm{Rama}}+U_{\mathrm{MJ}}
\label{e6}
\end{equation}
where $U_{\mathrm{HB}}$ is the hydrogen--bonding potential, 
$U_{\mathrm{Rama}}$ is the potential due to Ramachandran propensities, 
and $U_{\mathrm{MJ}}$ is the residue--specific potential 
based on MJ interactions. These terms are described below.

Hydrogen bonding for the backbone--backbone interactions is
modeled 
atomistically, but with simplifications appropriate to the otherwise
coarse--grained nature of our models. Specifically, we use ordinary
Coulomb interactions with OPLSAA
charges between the backbone CO and NH groups if
the O--H distance less than 2.5 $\mathrm{\AA}$. 
The cutoff was chosen as the distance after which dipole interactions are
significantly attenuated.
Following previous studies that suggest a dielectric constant of
2--5 inside a protein, we use a value of 3.
The use of physical charge and distance units in the hydrogen--bonding
interactions allows physical temperature units in the simulation (instead
of merely being in relation to the G$\bar{\mathrm{o}}$ well depth).

 Ramachandran propensities were included via the term
$U_{\mathrm{Rama}}$, which is based on a potential of mean force
obtained by calculating the distribution of $\Phi$--$\Psi$
dihedral angles in acetaldehyde--alanine--n--methylamide using
OPLSAA force field. This distribution was tabulated from a
Langevin dynamics simulation at 300~K using the GBSA implicit
solvent model in Tinker software package.

The construction of MJ--type interactions required some care.
Several variants of the original MJ interaction values have been
utilized in the literature (such as scaling the MJ interactions
energies, as well as shifting)\cite{mjmix,mj_schlik}
-- due to the fact that MJ values are based on folded protein 
data and are not directly applicable for unfolded states. 
We follow the suggestion 
of Jernigan and Bahar\cite{mjmix} to mix MJ values
of Table V and Table VI (numbering as in the original MJ paper,\cite{mj1} 
with updated values as in Ref~\cite{mj2}) so that the residue 
specific interactions are modeled as 
$x\times$(Table V)$+(1-x)\times$(Table VI).
We chose $x=0.05$ to ensure that the
residue--specific interactions are a significant perturbation
of the double G$\bar{\mathrm{o}}$ interactions.

To make the crystal structures (meta)stable, we ``titrated in''
double G$\bar{\mathrm{o}}$ interactions ($\epsilon$), 
until bistability
was observed at 300 K. Because, as described, hydrogen bonding 
introduces physical units into Model 2, the units of G$\bar{\mathrm{o}}$ well depth, 
$\epsilon$, are also physical.  We found that at $\epsilon /k=400$ K, 
both the structures remained (meta)stable.

\subsection{Model 3: Pure double G$\bar{\mathrm{o}}$ without energy symmetry}

Finally, to facilitate the generation of large path ensembles in both the 
Open--to--Closed and Closed--to--Open directions, we also constructed a third
model. The new model is designed to overcome the somewhat artifactual
over--stabilization of the Closed states in Models 1 and 2 (see results below
in Section~\ref{sec:result}).
In brief, our 8 $\mathrm{\AA}$ cutoff permits significantly more
contacts in the Closed state, implicitly but artificially mimicking 
the presence of ligands in Models 1 and 2. This implicit presence of ligands
interferes somewhat with our goal of modeling the ligand--free
opening and closing of the enzyme.

In Model 3, therefore, we attempt to
make the Open and Closed forms of adenylate kinase more
comparable in stability. We decrease the strength of G$\bar{\mathrm{o}}$
interactions specific to the Closed form to half of
G$\bar{\mathrm{o}}$ interactions ({\it i.e.}, we set $e_{\mathrm{scale}}=0.5$). 
Additionally, to focus on the effect of the reduced stability of the
Closed form with respect to the Open form, we use only
asymmetric double G$\bar{\mathrm{o}}$ interactions (and no H-bonding, 
Ramachandran, or MJ interactions). That is, we set
\begin{equation}
U_3=U_{\mathrm{G\bar{o}}} (e_{\mathrm{scale}}=0.5).
\label{e5}
\end{equation}

\section{Methods}
\label{sec:method}

\subsection{Dynamical Monte Carlo} 

We follow many precedents\cite{mcd1,mcd2,bin1} and use 
``dynamical'' MC for the dynamics of
our models. Such an approximation to physical dynamics is consistent
with our use of simplified models.
Specifically, we use the library--based Monte Carlo (LBMC) 
algorithm,\cite{lbmc}
to propagate the system in both brute--force simulations for
generating equilibrium ensembles and path
sampling simulations (discussed below in more detail).
For both equilibrium and path sampling,
the systems always evolve via ``natural'' LBMC dynamics,
and no artificial forces are used to 
direct conformational transitions, as explained below.

 Our LBMC simulations use the same trial moves described in our
earlier work.\cite{lbmc}. Namely, one flexible peptide plane in
the current configuration is swapped with one stored in the library,
and a $\Psi$ angle is also displaced by a small amount.

\subsection{Path sampling}

 In systems with rugged energy landscapes, such as proteins,
regular brute--force simulations are not efficient for studying
transitions. For this reason,
we use the statistically rigorous weighted--ensemble (WE)
path sampling method to generate path
ensembles of conformational transitions of adenylate kinase
between the Open and the Closed states.
This method preserves the natural system dynamics\cite{bin2} and
was used previously to study protein folding,\cite{we_fold1},
protein dimerization,\cite{we_ass}, and conformational
transitions of calmodulin using an alpha--carbon
model.\cite{bin}
Weighted ensemble studies the probability evolution of trajectories
in the configuration space using any underlying system dynamics.\cite{bin2}
In this work, we use the WE method to study transitions in  
more detailed models
to evaluate the effect of increasing chemical detail
on the transitions,
and to study questions of symmetry in forward and reverse directions.

The procedure to use weighted--ensemble path sampling
to study conformational transitions
is described in detail elsewhere,\cite{huber,bin,bin2} and here we 
describe our simple implementation briefly.
Prior to beginning the simulations,
we divide a one dimensional projection of the configurational
space ({\it i.e.}, the DRMS from the target structure in the 
present study) into
a number of bins. The DRMS is a ``progress coordinate''
or ``order parameter'' --
and is not necessarily the reaction coordinate. The progress
coordinate roughly keeps track of the progress to the
target state: the DRMS of structures close to the target
state is necessarily small. It is also possible to use multidimensional
or adaptively changing progress coordinates,\cite{bin,bin2}
but was not found necessary here.

In the weighted--ensemble method, an evolving set of trajectories and their
probabilities are tracked. Procedurally,
several independent trajectories are started in an initial
configuration and run for a short time interval $\tau$
(consisting of multiple simulation steps) with natural dynamics.
At the end of each $\tau$ interval, the progress of the trajectories
along the progress coordinate is noted 
({\it i.e.}, into which bin along the progress
coordinate each trajectory ends).
Once bins are tabulated after each $\tau$, trajectories
are ``split'' (replicated with divided probability) and combined.
This keeps the same number of trajectories in each occupied bin,
prunes low--weight trajectories and splits trajectories with high probability.
This splitting and combining of simulations is performed statistically as 
discussed elsewhere.\cite{huber,bin,bin2}
The probability remains normalized and all probability flows can be measured.

The full details of our WE simulations are as follows.
We employ LBMC to describe the natural system dynamics.
We utilize 25 bins between the two states, with
20 simulations (trajectories) in each occupied bin. The
end state is defined as being at a DRMS of 1.5 $\mathrm{\AA}$ from
the target crystal structure, a definition used in both directions.
Using this definition of the end state, we calculate the probability flux
of trajectories entering the target state at the end of each $\tau$.

It should be noted that value of the probability 
flux into either state -- and hence the rate -- depends upon
the precise definitions of the two states.
Although probability flows are good indicators of sampling quality,
precise numerical values of the rates are not of great interest in
our study of simple models with
Monte Carlo dynamics. In this work, we are interested in the
path ensembles and not the rates.

\section{Results}
\label{sec:result}

\subsection{Static analysis of conformational differences}
\label{conf}

 For reference, we first analyze the conformational differences between the 
two end--state static crystal structures to quantify the observed
differences in the Open and Closed configurations of Figures~\ref{fig:crystal}.
 Figure~\ref{fig:cont} shows
the $\alpha$--carbon distance difference map of pairs of residues in the
Open and the Closed crystal structures. A large positive value implies that
a pair is farther apart in the Open structure than in the
Closed structure, whereas a negative value is the opposite. By construction, 
the figure is symmetric about the diagonal.
A few features of the two structures easily emerge from
Figure~\ref{fig:cont}. The inter--residue
distances for most of the residue pairs are very similar in the two
crystal structures. The major differences are that the distances in the
Closed structure between
residues labeled LID (114--164) are closer to BD (31--60) and several
residues of CORE are smaller than the corresponding distances in the Open
structure. Thus, Figure~\ref{fig:cont} quantifies Figure~\ref{fig:crystal}.

 From Figures~\ref{fig:crystal} and \ref{fig:cont} it is clear
that the structural change that characterizes the transition between
the Open and the Closed structure is fairly straightforward: the
LID and the BD close, and the rest of the protein
remains fairly unchanged. Following Figures~\ref{fig:crystal} and 
\ref{fig:cont}, for the path sampling studies presented shortly, we monitor 
inter--residue distances
between two pairs of residues: residues 56 (GLY) and 163 (THR), which
report on the BD--LID proximity, as well as
residues 15 (THR) and 132 (VAL), which report on the CORE--LID proximity. 
In the Closed structure, $d_{56,163}^{\mathrm{c}} = 4.9 \mathrm{\AA}$ and 
$d_{15,132}^{\mathrm{c}} = 6.3 \mathrm{\AA}$.
On the other hand, in the Open structure,
$d_{56,163}^{\mathrm{o}} = 23.6 \mathrm{\AA}$ and 
$d_{15,132}^{\mathrm{o}} = 17.8 \mathrm{\AA}$.
Thus, the relation between the CORE and LID is monitored,
along with that of the LID and BD.

\subsection{Brute force equilibrium sampling}

 In order to demarcate the native basins in our analysis of transitions,
we first study equilibrium ensembles for the Open and the
Closed states of adenylate kinase. Put another way, we want to quantify the
size of native--basin fluctuations in our models. Further, we determine
whether transition paths can be obtained without the aid
of path sampling.

 We quantify fluctuations in the equilibrium ensembles in
the two basins by using DRMS from the respective crystal structures.
Figure~\ref{fig:DRMS_go} (a) shows two sets of DRMS traces for Model 1 
for a simulation started from the Open structure: DRMS--from--Open
(black line) and DRMS--from--Closed (blue line). Similarly,
Figure~\ref{fig:DRMS_go} (b) shows two sets of DRMS traces for Model 1 
for a simulation started from the Closed structure: DRMS--from--Closed
(black line) and DRMS--from--Open (blue line). Thus, in each panel,
the black line represents DRMS from the starting structure,
whereas the blue line represents DRMS from the opposing structure.

 A comparison of the two panels of Figure~\ref{fig:DRMS_go}
shows that the simulation started from the Open structure shows significantly
more fluctuations than the simulation started from the Closed structure.
Furthermore, the fluctuations drive the simulation started from the Open
structure closer to the Closed structure than vice versa. For example,
Figure~\ref{fig:DRMS_go} (a) shows that the simulation started from the
Open structure gets to within 3 $\mathrm{\AA}$ of the Closed structure at
approximately 70 million MC steps. On the other hand, the simulation
started from the Closed structure (Figure~\ref{fig:DRMS_go} (b)) remains
farther from the Open structure.

 Most importantly, 
 neither simulation show a transition to the opposing structure.
The DRMS from the opposing structure for a particular simulation is 
always significantly larger than
DRMS values from the starting structure for the other simulation. To
elaborate, let us consider the DRMS--vs--Closed structure for the
simulation started from the Closed structure
(black line in Figure~\ref{fig:DRMS_go} (b)). The fluctuations in 
DRMS remain less than
1.5 $\mathrm{\AA}$ in the native basin for the Closed structure.
Comparatively, the largest fluctuations in the simulations started
from the Open structure bring it only within at most 3 $\mathrm{\AA}$
of the Closed structure (blue line in Figure~\ref{fig:DRMS_go} (a)).
That is, the opposing native basin is never reached.

 We mention that all the DRMS values plotted in Figures~\ref{fig:DRMS_go} (a)
and (b) are based on the first 200 residues. This is because
the 14 tail residues, which form a helical segment, are very
flexible and the helix unravels in either structure at a much lower
temperature than the stable part of the protein. Thus, Figure~\ref{fig:DRMS_go}
focuses on the rest of the protein. Additionally,
although we show results for $T=0.7$ here, simulations at lower
temperatures also give qualitatively similar results.

 We perform an analogous fluctuation analysis for Model 2 which incorporates 
backbone hydrogen bonding interactions, Ramachandran propensities, and some
residue specificity via MJ--type interactions.
Figure~\ref{fig:DRMS_mj} (a) shows the DRMS (of the first 200 residues) 
from the Open (black line) and Closed (blue line) structures for a simulation
started from the Open structure. Similarly,
Figure~\ref{fig:DRMS_mj} (b) shows the DRMS traces for a simulation
started from the Closed structure. 
Again, we observe very similar results as for Model 1: the fluctuations
in the Open ensemble are larger than in the Closed ensemble, and no
transition to the opposing structure is obtained in either simulation.

\subsection{Path sampling: Models 1 and 2}

 Due to the inability of brute--force simulations to show transitions,
we use weighted--ensemble path sampling to generate an ensemble
of transition pathways with the aim of assessing path heterogeneity.
In particular, we examine transitions in both directions
for all the three models.

\subsubsection{Transition from Open to Closed State}

 We first check whether our path sampling is sufficient
by monitoring the flux into the target state.
Figure~\ref{fig:rate_1ake_4ake} plots the WE results for probability 
fluxes obtained into the Closed state for both Models 1 and 2. 
The ``time'' axis is merely the number of $\tau$ intervals
(where one interval contains 2000 LBMC steps). 
In both models, the fluxes reach linear regimes
indicating that the observed transitions are not merely due
to initial fast trajectories and the path ensemble is
appropriately sampled.

The sensitivity to the models is also apparent in the fluxes
shown in Figure~\ref{fig:rate_1ake_4ake}:
Model 2 (which includes hydrogen--bonding, Ramachandran propensities, and
MJ--type residue specific interactions) has a smaller flux into
the Closed state than Model 1. Residue--specific interactions
are expected to roughen the energy landscape, consistent with
the observed slowing of transition dynamics. However, the possible
change in the Open state basin stability due to addition
of these interactions is convoluted with the roughening of
the landscape.

We further study the path ensemble by examining individual trajectories.
Figure~\ref{DRMS_we_go} shows, for Model 1, the DRMS
from the Closed structure for four typical
transitions started in the Open state as a function of time 
(total number of LBMC steps) obtained via WE path sampling. 
In contrast with the brute--force simulation in Figure~\ref{fig:DRMS_go} (a), 
each trajectory in Figure~\ref{DRMS_we_go} gets to the 
Closed state (defined to be within
a DRMS of 1.5 $\mathrm{\AA}$ from the Closed structure).
Although the trajectories arrive at the target state with
different weights, the ones shown in the figures above are obtained after
a simple resampling procedure,\cite{liu} and, thus, represent
trajectories that arrive with relatively large probabilities.
Resampling is a statistically rigorous procedure to
prune an ensemble.\cite{liu} In our resampling scheme, 
a trajectory arriving at
the target with weight $w$ is kept
with a probability $w/w_{\mathrm{max}}$.

For trajectories that begin transitions at larger times (such as
Trajectory 4 in Figure~\ref{DRMS_we_go}), a significant amount 
of time is spent in regions
with large DRMS values from the Closed structure. 
Thus, in Figure~\ref{fig1} and beyond, we do not show the ``dwell time''
in the Open state.

To analyze the order of domain closing, and, in particular, to study
possible heterogeneity in the path ensemble, we study the
four trajectories of Figure~\ref{DRMS_we_go} in more detail.
Figure~\ref{fig1} plots the projection of the above four trajectories
onto the  $d_{15,132}$ and $d_{56,163}$ plane (see Section~\ref{conf}) 
obtained via WE for
Open--to--Closed transition. The filled circle shows the Closed x--ray 
structure,
whereas the filled diamond is for the Open x--ray structure. The corresponding
open circles and diamonds are representative of fluctuations in the ensembles
of Closed and Open structures, respectively. 
The relatively larger spread of structures in the
Open ensemble compared to Closed reflects the
larger Open--state fluctuations depicted previously
in Figure~\ref{fig:DRMS_go}. 
The four different colored lines show the four trajectories of
Figure~\ref{DRMS_we_go}, without the ``dwell time'' in the Open ensemble.

In all the trajectories, the transition through the region
with values of $d_{56,163}$ (BD--LID distance) intermediate between the two
ensembles is fairly rapid. The flexible LID undergoes large
fluctuations in the Open state, and the transition to the
Closed state is typically accomplished via the BD snapping
closed on a much smaller timescale. 

 Despite the relatively fast closing of the BD for the trajectories
in Figure~\ref{fig1}, the exact transition
paths traced by the four trajectories are significantly different.
For Trajectory 1, the BD shuts after the flexible
LID gets close to the CORE. For the following discussion,
we call this pathway as Open--LID--BD--Closed (first the LID relaxes,
and then the BD shuts close). On the other hand, Trajectory 3
shows a dramatically different behavior: the BD
snaps shut before the flexible LID gets closer to the CORE (this
pathway is labeled as Open--BD--LID--Closed). The other two
trajectories are somewhere in between the two extremes.

To quantify heterogeneity in the path ensembles, we compare
the ratio of trajectories in the two transition pathways.
Specifically, we define a trajectory to follow the Open--LID--BD--Closed
(lower right) pathway if it first visits the region 
$d_{56,163}<10.0$ $\mathrm{\AA}$ after last leaving the rectangular
Open--state region defined by $d_{56,163}>10.0$ $\mathrm{\AA}$ and
$d_{15,132}>10.0$ $\mathrm{\AA}$.
On the other hand, a trajectory follows 
the Open--BD--LID--Closed (upper left) pathway if it first visits the region 
$d_{56,163}>10.0$ $\mathrm{\AA}$ after last leaving the above Open--state
rectangular region. We find that for Open--to--Closed transition using
Model 1, approximately 60$\%$ of the resampled trajectories 
follow Open--BD--LID--Closed pathway (akin to Trajectory 3 in
Figure~\ref{fig1}). The remaining 40$\%$ follow the
Open--LID--BD--Closed pathway.

 Further, we look at a few intermediate structures for these
two pathways. Figure~\ref{fig:1c} shows four intermediates
along Trajectory 3 of Figure~\ref{fig1}.
The BD and LID domains near one another
before the LID closes. On the other hand,
Figure~\ref{fig:1d} shows four intermediates along Trajectory 
1 in Figure~\ref{fig1}. The closing of the LID,
 followed by snapping shut of the BD
is clearly visible in the figure. As both Figures~\ref{fig:1c} and
\ref{fig:1d} show, the rest of the protein ({\it i.e.}, the CORE
region) maintains a stable shape during the transformation.

To determine the sensitivity of the path ensemble to the model, we
similarly analyzed results from Model 2
(which includes hydrogen bonding, Ramachandran propensities, 
and a level of residue specificity).
A similar qualitative picture is obtained for Model 2.
Figure~\ref{fig3}
plots three of the resampled trajectories
from the Open to the Closed structure using Model 2. 
The ``dwell times'' in the Open state have been omitted for clarity.
Again, the symbols have the same meaning as in Figure~\ref{fig1} 
(except that open symbols represent the fluctuations obtained using Model 2).
The transition from the Open--to--Closed structure primarily
takes place by the BD snapping closed to the
LID on a much shorter time scale. Depending upon the
relative positions of the LID and CORE, the
completion of the transition requires further adjustment of
the LID relative to the CORE.
The ratio of paths in the two pathways is the same as that for
Model 1.

\subsubsection{Transition from Closed to Open State}

 We also studied ``reverse'' transitions -- from the Closed
to the Open state. Figure~\ref{fig:rate_4ake_1ake} shows the flux 
into C as a
function of ``time'' for both Models 1 and 2.
Compared to Figure~\ref{fig:rate_1ake_4ake} 
for the transition
from the Open to the Closed state, the flux into state B is
several orders of magnitude lower. This observation mirrors the previously
described larger fluctuations in the Open state ensemble.
Flux into the Open state for Model 2 with
residue specific chemistry is higher than for Model 1,
despite the expected roughening of the energy landscape.
This necessarily reflects a free energy shift, suggesting MJ
interactions de--stabilize the Closed state compared to a pure
double--G$\bar{\mathrm{o}}$ model.
Such a shift seems appropriate given that 
we do not model ligands which implicitly lead to more
contacts in the Closed state and consequent over--stabilization in the
G$\bar{\mathrm{o}}$ model.

 For Closed--to--Open transition using either model, 
we obtain pathways which mirror
the Open--to--Closed transition: the LID
fluctuates in the Closed state, and this is followed by
the BD snapping open on a relatively
fast time scale. For both Models 1 and 2, successful trajectories
appear to follow only the Closed--LID--BD--Open pathway for 
Closed--to--Open transition
(reverse order of the Open--BD--LID--Closed pathway in the Open--to--Closed
transition direction). The absence of symmetry is surprising
given our recent formal demonstration,\cite{symm} and there seem to be two
possible reasons.
First, the transients for Closed--BD--LID--Open
pathway are long--lived. Lengthy transients are consistent with the low
reverse reaction rates, shown in Figure~\ref{fig:rate_4ake_1ake},
for both Models 1 and 2.
Second, our state definitions may be flawed as discussed in Section~\ref{sym}.

To clarify the issue of the symmetry of path ensembles between
forward and reverse directions, we constructed and path sampled Model 3.

\subsection{Path ensemble symmetry analysis in Model 3}

 The slow Closed--to--Open transitions indicates that, for Models 1 and 2,
the free energy of the Closed structure is significantly
lower than that of the Open structure. As discussed in Section 2.3, this
suggested the use of Model 3, which decreases in magnitude 
the favorable energy for contacts present only in the Closed state.
That is, Model 3 reduces the
free energy asymmetry between the Open/Closed states.

Model 3 thereby facilitates study of the symmetry between forward and
reverse transitions.
As shown in Figure~\ref{fig:rate_m3}, although the flux in the
Open--to--Closed direction in Model 3 is higher than in the Closed--to--Open
direction, the difference between the fluxes in the two directions
is much less than that for Models 1 and 2. The increased
Closed--to--Open rate implies that the relative stability of
the Closed state is reduced compared to
Models 1 and 2.
Importantly, the relatively linear behavior of fluxes in both the directions
implies our path sampling is sufficient -- well beyond transients.

 For Model 3, we examine the same classification of pathways as above.
Both paths are frequently observed in both directions.
In Figure~\ref{fig:avgpath_m3}, we show the ratio of probabilities of
the two paths as a functions of simulation time in the two directions.
Values in each window are averaged over 500 $\tau$ increments.
The results for the Open--to--Closed direction (diamonds) are shown for a
single simulation, whereas the Closed--to--Open transitions (circles) are shown
for 6 independent simulations.  Despite
large fluctuations, the ratios of paths in the two directions are similar.
We discuss the issue of path symmetry further, below.

\section{Discussion}
\label{disc}

\subsection{Models}
\label{mdl}

An important issue in any coarse--grained study 
is the sensitivity of the results to the particular model(s) used. 
To address this point, we used three different semi--atomistic models of
adenylate kinase.
For the models used, we find that the transition pathways are
not significantly affected by the models we used. In particular,
we find two dominant pathways (Open--LID--BD--Closed and 
Open--BD--LID--Closed) that occur in all the models.
Although the rates vary considerably among models, we do not expect
realistic kinetics in simplified models.

Our choice of models was governed by the basic requirement of obtaining
full path sampling of conformational transitions -- in order to study
path ensembles, heterogeneity, and symmetry.
Two of the models are based purely on structure (G$\bar{\mathrm{o}}$ 
model) and the other (Model 2) includes some level of residue specificity via
Miyajawa--Jernigan interactions, as well as hydrogen bonding
energies and Ramachandran propensities. In Model 2, the chemical
energy terms are
significant perturbations to the G$\bar{\mathrm{o}}$ interactions.
(as quantified by MJ interactions between residues). 
This model is designed to be able
to capture a minimum level of biochemistry. However,
Model 2 still requires significant G$\bar{\mathrm{o}}$--type interactions to
stabilize the two physical states. In the future, we plan to
utilize more detailed and explicit side chain--side chain and
side chain--backbone interactions to reduce the dependence on G$\bar{\mathrm{o}}$--type
interactions.

 Another limitation is that we did not consider the ligand in our path sampling
simulations. The inclusion of ligand could influence
the observed pathways significantly. We have plans for modeling ligand 
via ``mixed models''
that include all--atom ligands and binding sites, with a coarse--grained
picture for the rest of the protein. Such an explicit inclusion
of ligands, with the corresponding degrees of freedom of the
unbound ligands in the Open form should reduce the dependence on
arbitrary G$\bar{\mathrm{o}}$ interactions.
A study with explicit ligands could require a 
higher dimensional progress coordinates to use in weighted ensemble simulations:
one coordinate for protein structure (as is done in this work),
and a second (or further) coordinates for the distance between ligands and 
the protein. Note that weighted ensemble can mix 
real-- and configurational--space coordinates: it was originally
designed for binding studies.\cite{huber}

\subsection{Path symmetry}
\label{sym}

 Recently, we investigate the conditions when there should be
symmetry -- {\it i.e.}, when pathways in the forward and the reverse
directions occur with the same ratio.\cite{symm}
We show that exact symmetry will hold when a specific
(equilibrium--based) steady state is enforced. Approximate symmetry
is expected if the initial and final states are well--defined
physical basins lacking slow internal timescales, so that trajectories
emerging from a state ``forget'' the path by which they entered.
Figure~\ref{fig:avgpath_m3} suggests that the ratio of the two different
pathways in the two directions is very similar for Model 3, which was
fully path sampled in both directions. 

Such a symmetry is clearly absent from our results 
(even after accounting for statistical fluctuations) in Models 1 and 2.
Although we observed transitions in both the directions for all
the models, Closed--to--Open transitions in all the models (especially
in Models 1 and 2) are harder to obtain.
In particular, the Closed--BD--LID--Open pathway is not observed
in our simulations for Models 1 and 2. This indicates a lack of
the correct steady state
for these models in the Closed--to--Open direction and/or
insufficiently well--defined states. It is unlikely that
the highly flexible Open state is a good physical basin.
We are currently working on developing WE path sampling
methods that allow steady states to be sampled 
directly and efficiently.\cite{ss} Related steady--state methods
are already available.\cite{dinner1,dinner2,van1}

\subsection{CPU time and efficiency}

One of the basic goals of this work was to determine the level of detail
we can include in a model, while still allowing for full sampling of the
path ensemble. Thus, we now discuss the computational effort that was 
required. All simulations were performed on single 3 GHz Intel processors.
The results shown for Model 1 in the Open--to--Closed direction took
approximately one week of single CPU time.
More simulation was performed in the Closed--to--Open
direction, requiring 3-4 weeks of single CPU time. The results
for Model 2 were obtained using approximately the same time as Model 1.
For Model 3, the Closed--to--Open transition was
not much harder to obtain than the Open--to--Closed transition, and
a simulation in each direction required approximately two weeks 
of single CPU time.
Due to the low CPU usage for obtaining path ensembles for the models used
here, obtaining path ensembles of better models using WE is
possible. See Section~\ref{mdl}.

It is not hard to estimate the efficiency of WE simulation compared
to brute--force.
The transition rates determined from WE simulations
indicate the time required
for brute--force simulations to achieve transitions and hence permit estimates
of efficiency. For example, the rate obtained
for Closed--to--Open transition for Model 3 is 2.5$\times 10^{-6}/\tau$.
Thus, one brute--force transition can be estimated to require 
the reciprocal amount of
time. Since 2000$\tau$ require approximately one week of computing,
BF is estimated to take approximately 4 years for a single transition. 
In contrast WE yielded 50 transitions after resemapling ({\it i.e.},
50 transitions with equal weights), in  about two weeks of single--processor
computing. (Before resampling, there were about 3000 WE transitions for
each simulation).
WE is thus significantly
more efficient than BF. For transitions in the other direction and/or
other models, a qualitatively similar picture for efficiency emerges.

\section{Conclusions}
\label{cnc}

 We applied weighted ensemble (WE) path sampling to generate ensembles 
for conformational transition between Open (apo) and Closed (holo)
forms of adenylate kinase using semi--atomistic models of the protein.
No additional driving force was used to enable the transitions.
We showed that conformational transitions
in both directions are possible for such models via WE. In contrast,
brute--force simulations are vastly inefficient.
Given the relatively small computational effort required for
observing transitions using WE, more detailed models can be used for 
full path sampling.  In the future, models with further reduced dependence on 
G$\bar{\mathrm{o}}$--type interactions are needed, along with 
ligand modeling, to study the specific enzyme 
biochemistry -- and path sampling of such models appears possible.

All the models show significant hereteogeneity in the transition pathways. 
In particular, two dominant pathways observed are characterized by
the order in which the flexible lid and the AMP binding domains close.
Although the rates obtained (in terms
of Monte Carlo steps) varied significantly depending upon the
model used, similar dominant pathways are obtained across the
models. We further showed in the Appendix the formally exact result
that the transition paths must be symmetric in the two directions in the
(equilibrium--based) steady states. The model that allows significant
transitions in both the directions shows an approximate symmetry which
appears to be consistent with conditions on the symmetry rule.

{\bf Acknowledgments:} We thank Dr. Bin Zhang and  Prof. David Jasnow
for helpful discussions. This work was supported by the NIH (Grant GM070987)
and the NSF (Grant MCB--0643456).

\clearpage

\bfig
\begin{tabular}{c}
\resizebox{2.5in}{!}{\includegraphics{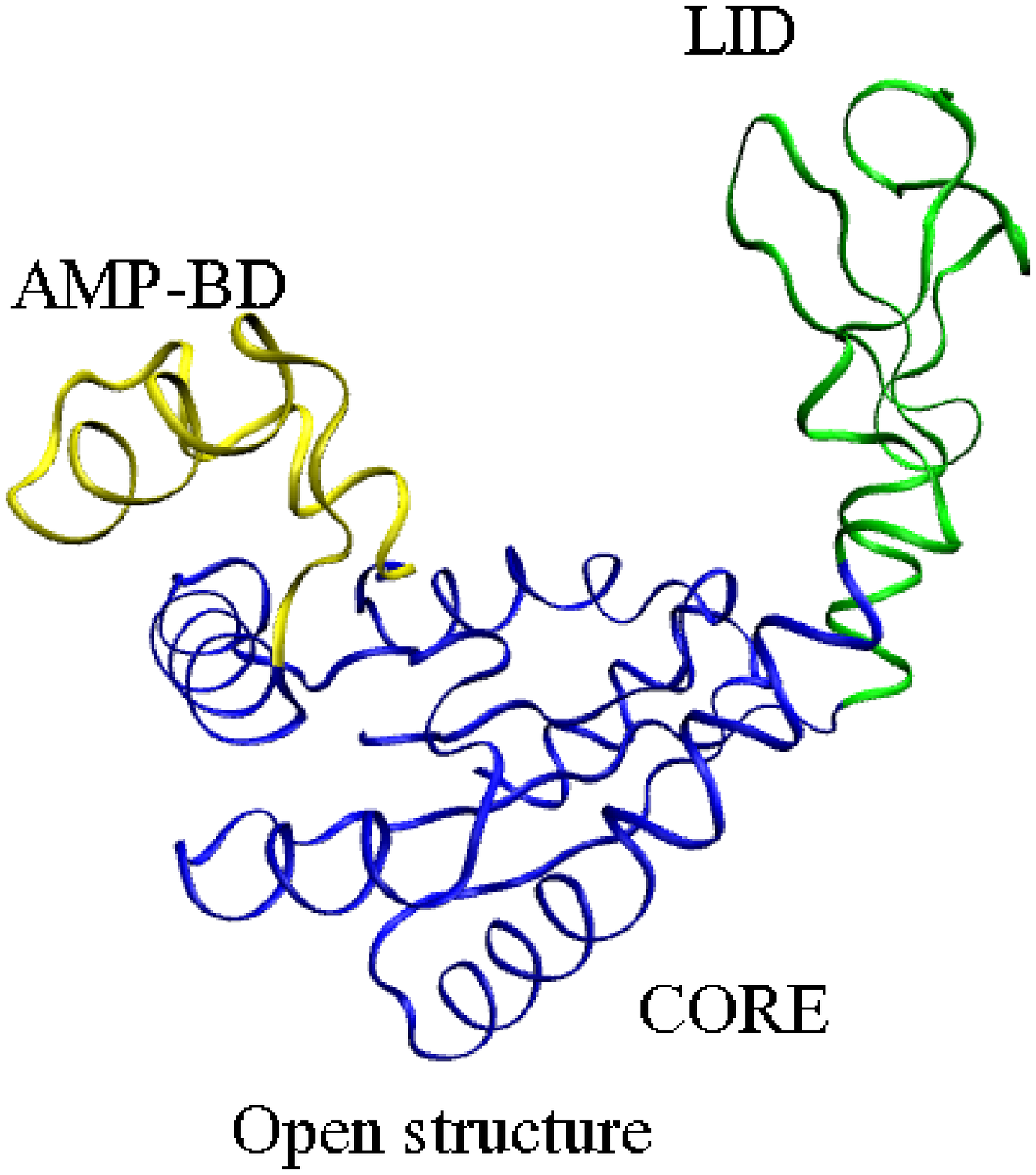}} \\
\resizebox{2in}{!}{\includegraphics{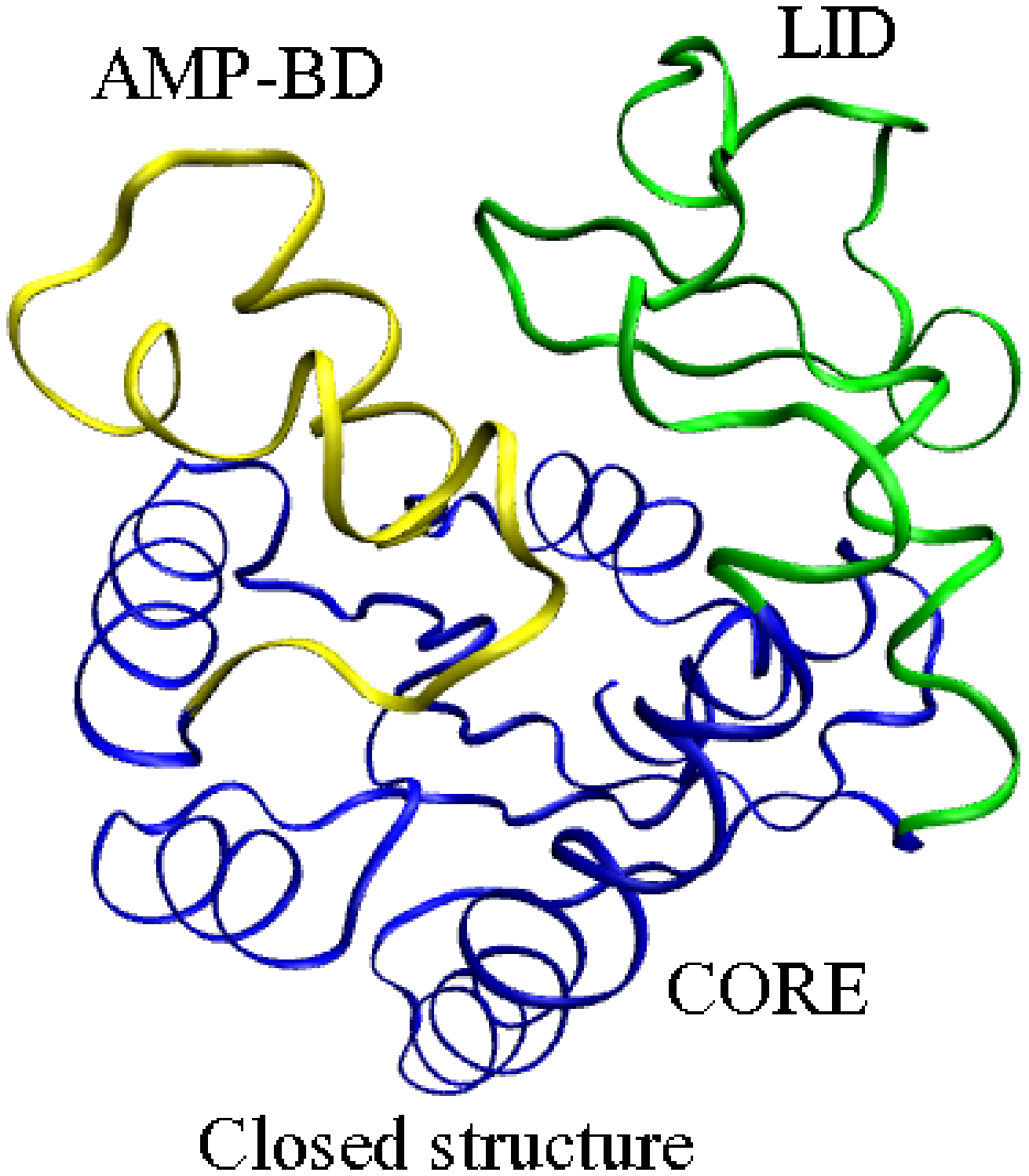}} \\
\end{tabular}
\caption{Crystal structures of adenylate kinase. The blue segment represents
the core of the protein (CORE), the yellow segment is the AMP
binding domain (BD), and the green segments is the flexible lid (LID). The top
figure is the Apo (or Open) form, and the bottom figure is holo (Closed)
form.}
\label{fig:crystal}
\efig

\clearpage

\bfig
\resizebox{3.5in}{!}{\includegraphics{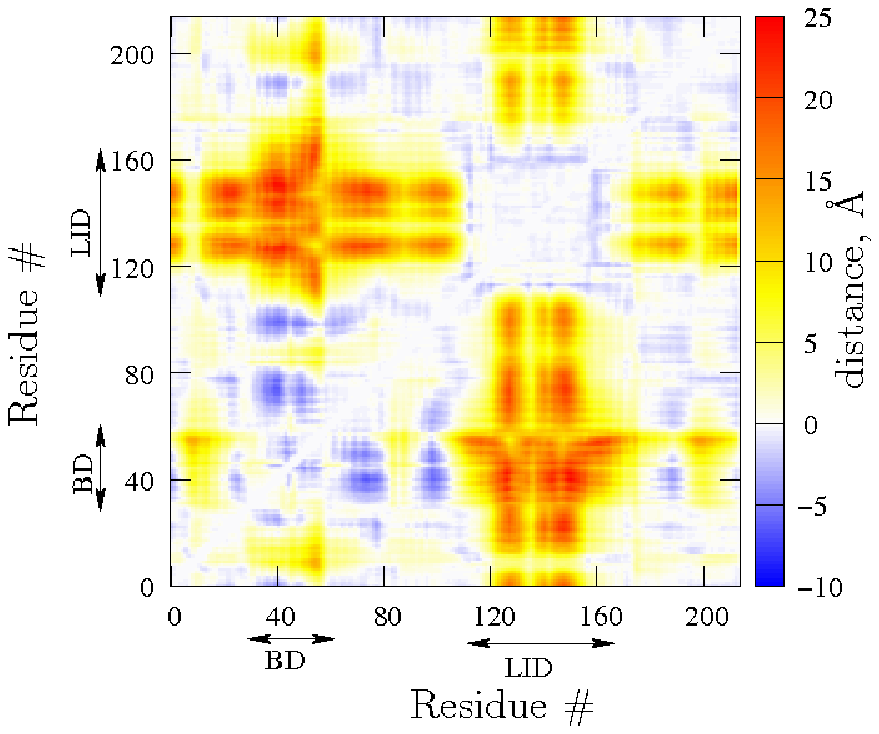}}
\caption{Map of the difference in the inter--residue distances of the
Open from the Closed crystal structure.
The map is, by definition, symmetric about the diagonal. The white
space on the diagonal just implies that we do not calculate
distances between residue pairs less than 2 residues apart along
the chain. The residues corresponding to the LID and BD are labeled and
the unlabeled residues form the CORE.}
\label{fig:cont}
\efig

\clearpage

\bfig
\begin{tabular}{c}
\resizebox{3.5in}{!}{\includegraphics{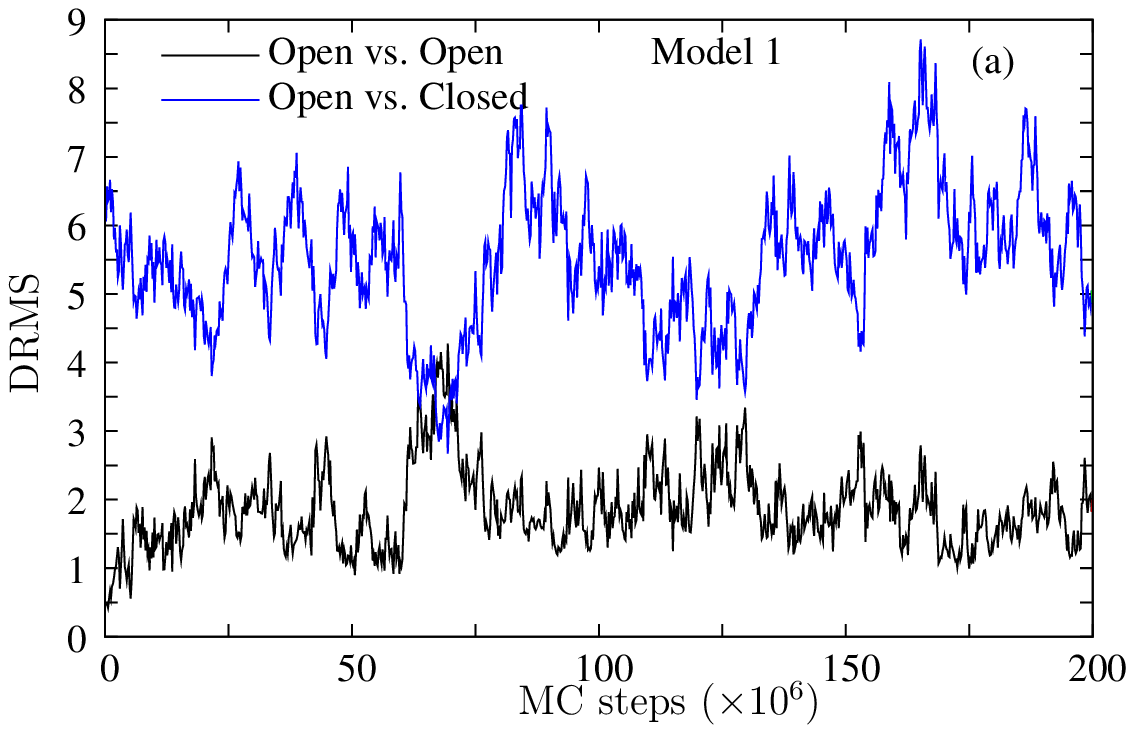}} \\
\resizebox{3.5in}{!}{\includegraphics{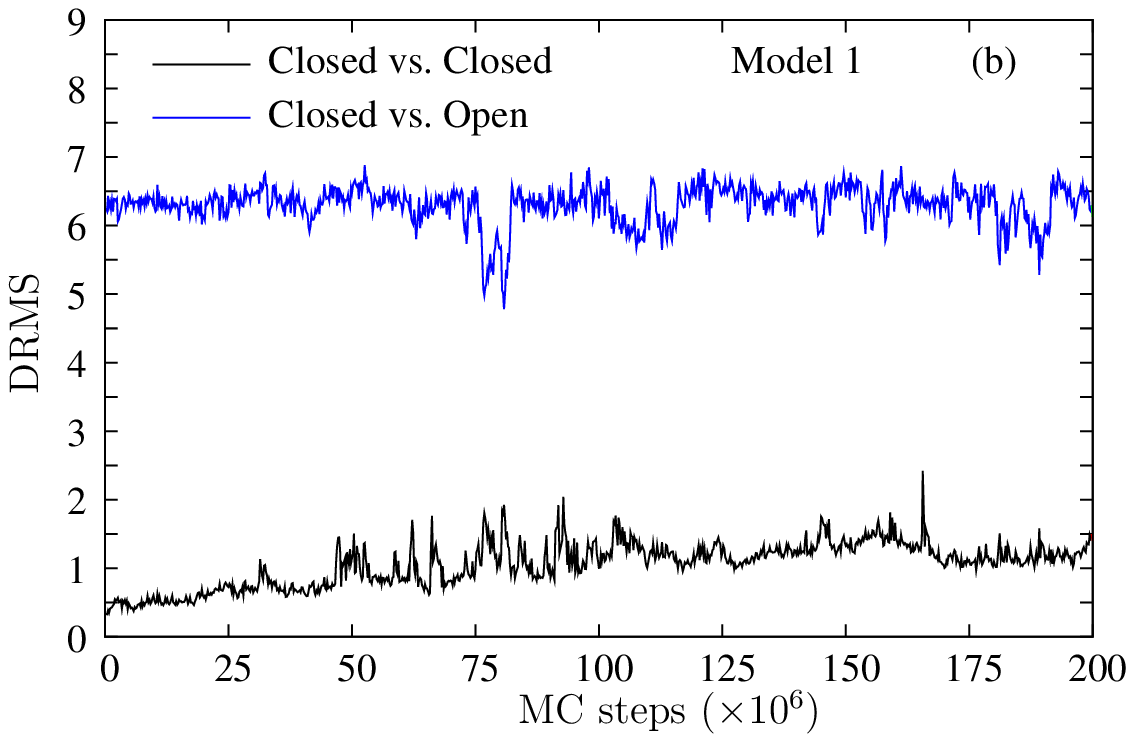}} \\
\end{tabular}
\caption{Stability of the native basins in Model 1. The DRMS in 
Model 1 (pure G$\bar{\mathrm{o}}$) 
is shown for two simulations (a) starting
from the Open structure and (b) starting from the Closed structure.
For each simulation, we show the DRMS from the starting structure
(black line) and the opposing structure (blue line). Neither simulation
show a transition to the opposing structure, but the scales of fluctuations
are very different.}
\label{fig:DRMS_go}
\efig

\clearpage

\bfig
\begin{tabular}{c}
\resizebox{3.5in}{!}{\includegraphics{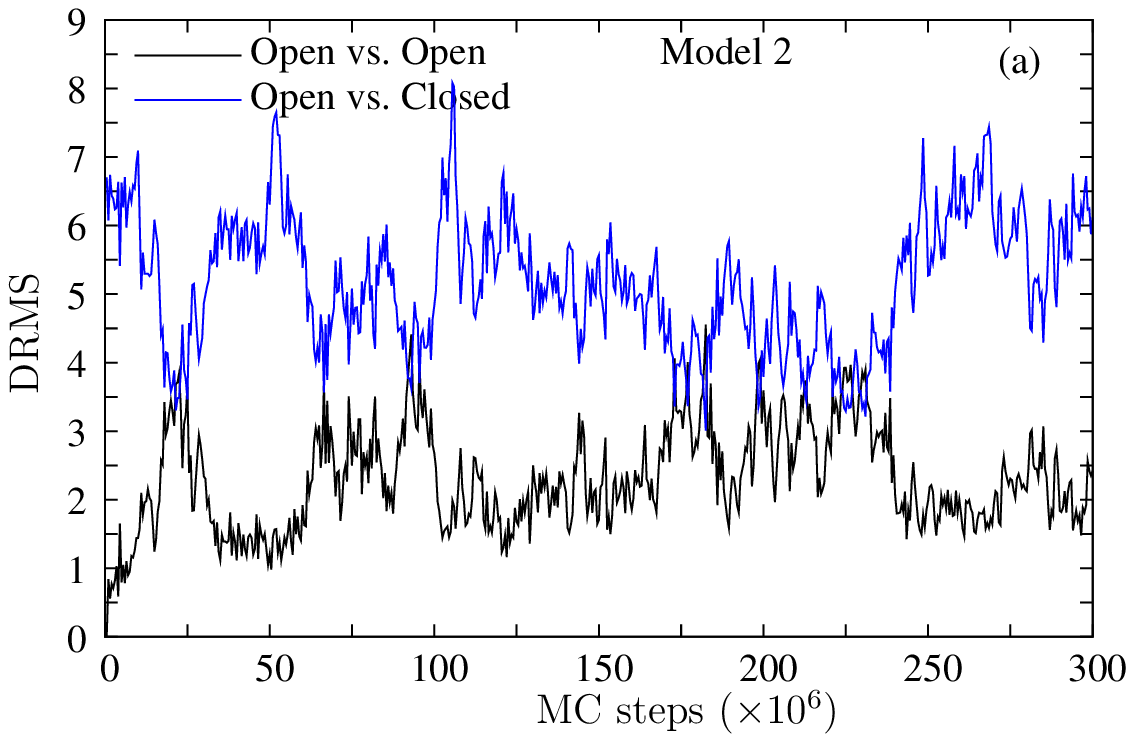}} \\
\resizebox{3.5in}{!}{\includegraphics{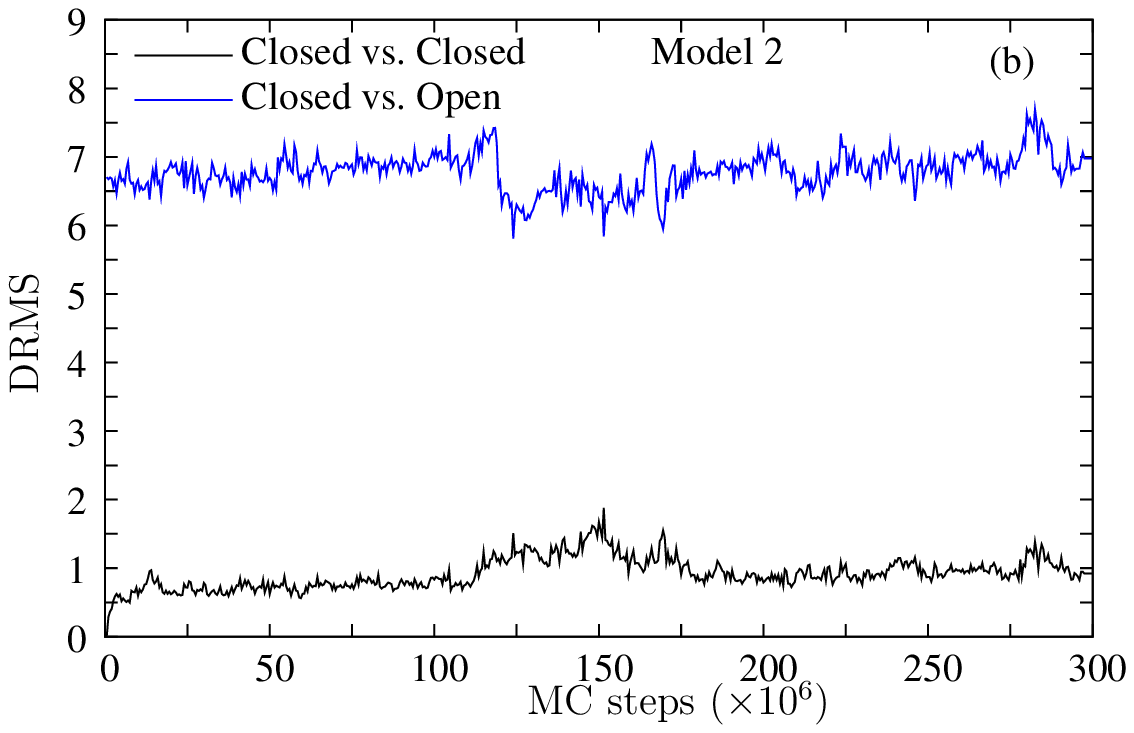}} \\
\end{tabular}
\caption{Stability of the native basins in Model 2. 
The DRMS in Model 2 (G$\bar{\mathrm{o}}$, H-bonding, Ramachandran propensities,
and MJ--type residue specific interactions) is shown for two simulations 
(a) starting
from the Open structure and (b) starting from the Closed structure.
For each simulation, we show the DRMS from the starting structure
(black line) and the opposing structure (blue line). Neither simulation
show a transition to the opposing structure, but the scales of fluctuations
are very different.}
\label{fig:DRMS_mj}
\efig

\clearpage

\bfig
\resizebox{3.5in}{!}{\includegraphics{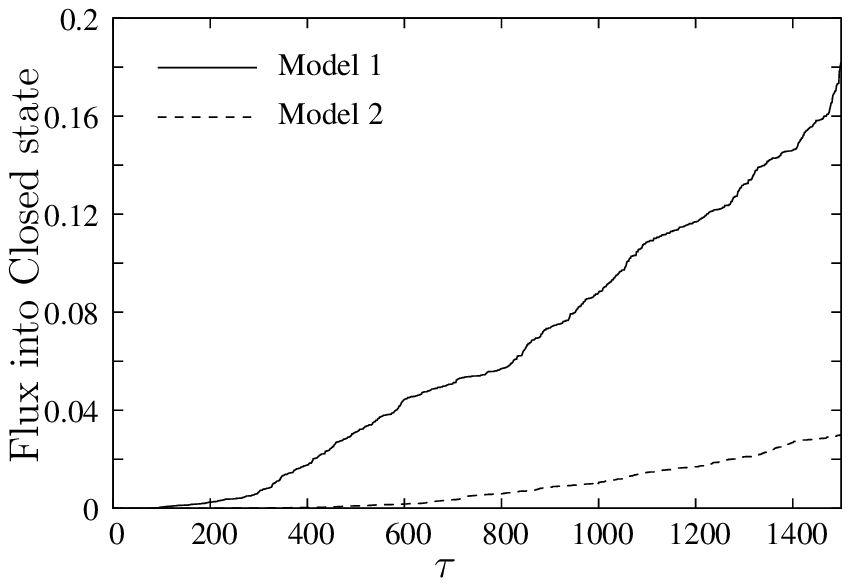}}
\caption{Comparison of the probability fluxes into the Closed state 
for two models.
The fluxes from a pure double--G$\bar{\mathrm{o}}$ system (Model 1, solid
line) and a system with considerable chemical specificity (Model 2,
dashed line) are plotted
as functions of WE time intervals (each $\tau$ interval is 2000 LBMC steps). 
In both cases, the fluxes reach approximately linear regimes,
suggesting transient effects have attenuated.}
\label{fig:rate_1ake_4ake}
\efig

\clearpage

\bfig
\resizebox{3.5in}{!}{\includegraphics{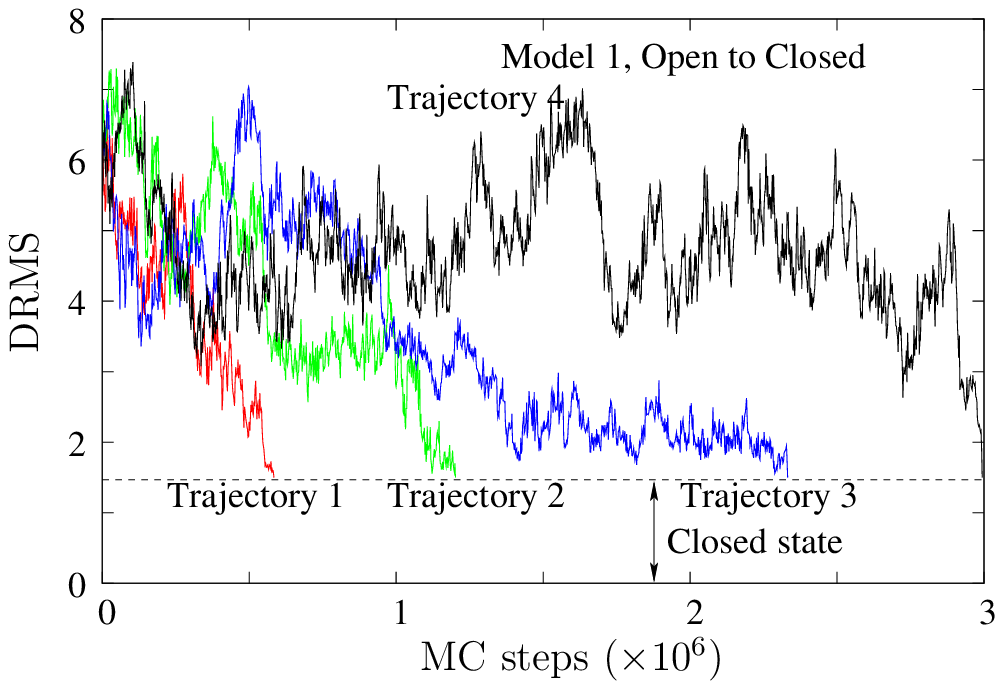}}
\caption{Open--to--Closed transitions observed for Model 1.
Four typical DRMS traces are shown, measured from the Closed crystal structure.
The Closed basin is delimited by a DRMS of 
1.5 $\mathrm{\AA}$ from the Closed crystal structure.}
\label{DRMS_we_go}
\efig

\clearpage

\bfig
\resizebox{3.5in}{!}{\includegraphics{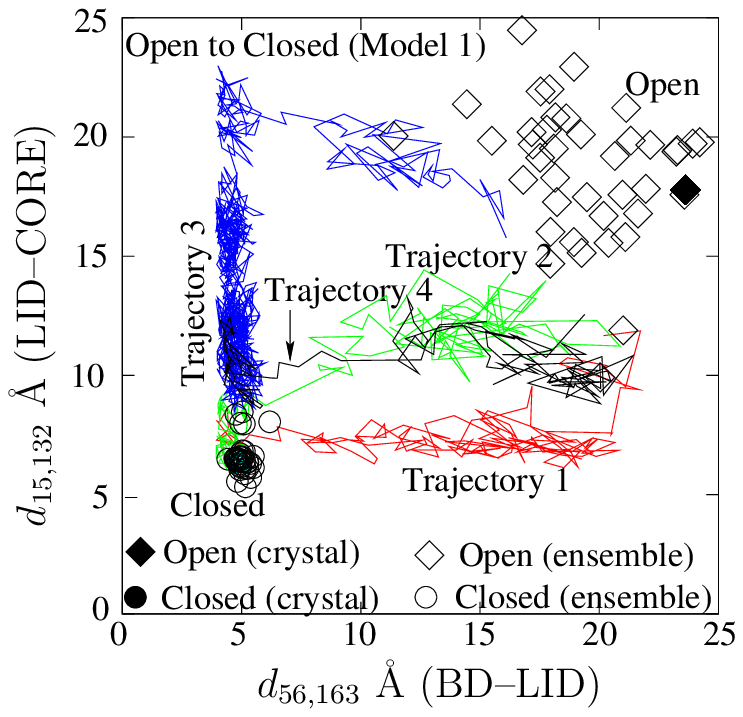}}
\caption{Path heterogeneity in Model 1 transitions. Four typical trajectories
for the Open--to--Closed transition (Figure~\ref{DRMS_we_go} are
shown via distances between
the CORE and LID (ordinate) and between the BD and
the LID (abscissa). The ``dwell'' times for the trajectories in
the Open state excluded for clarity.}
\label{fig1}
\efig

\clearpage

\bfig
\begin{tabular}{cc}
\resizebox{2.0in}{!}{\includegraphics{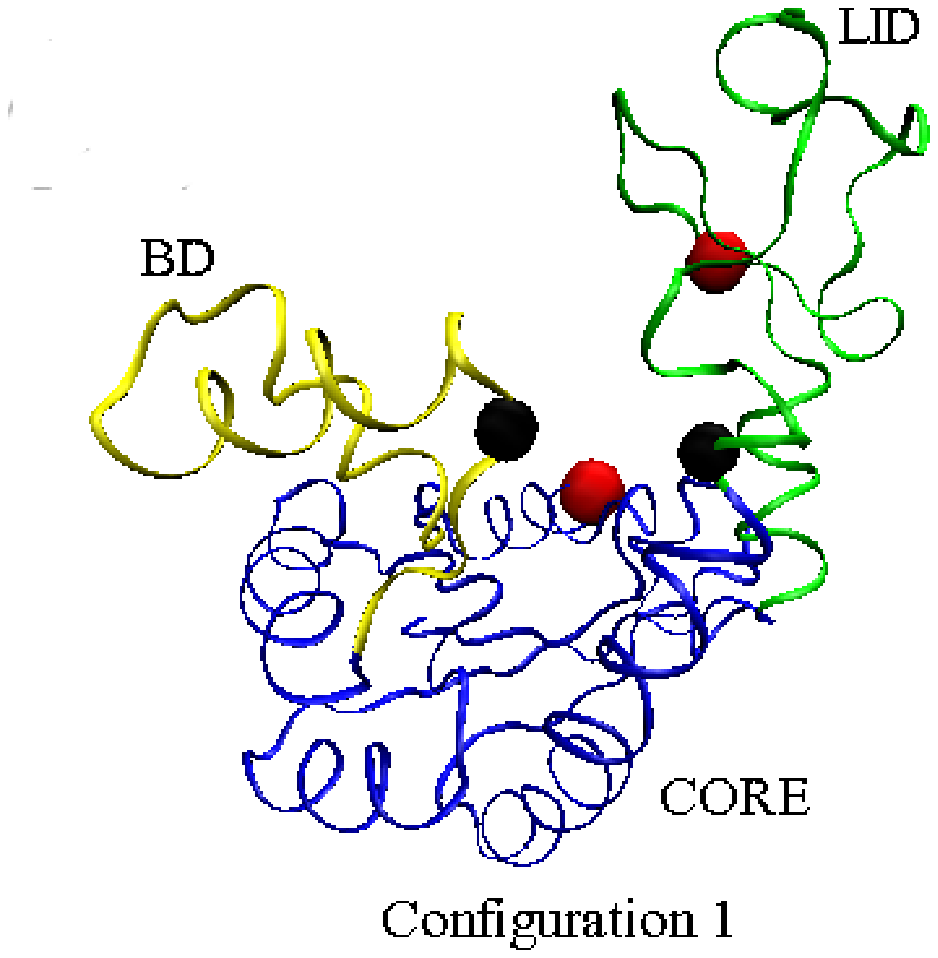}} &
\resizebox{2.0in}{!}{\includegraphics{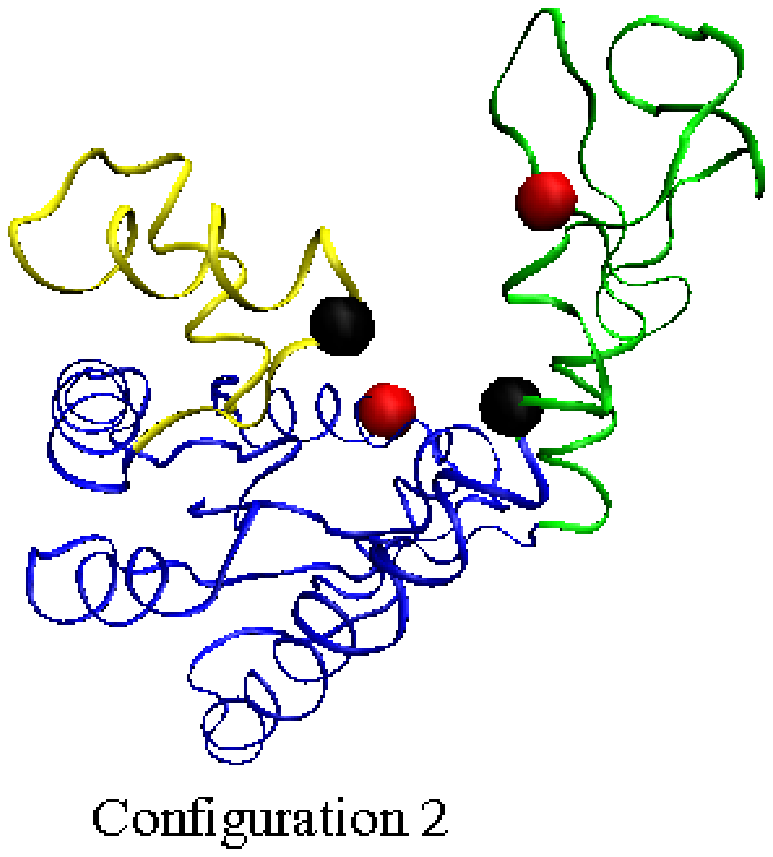}} \\
\resizebox{2.0in}{!}{\includegraphics{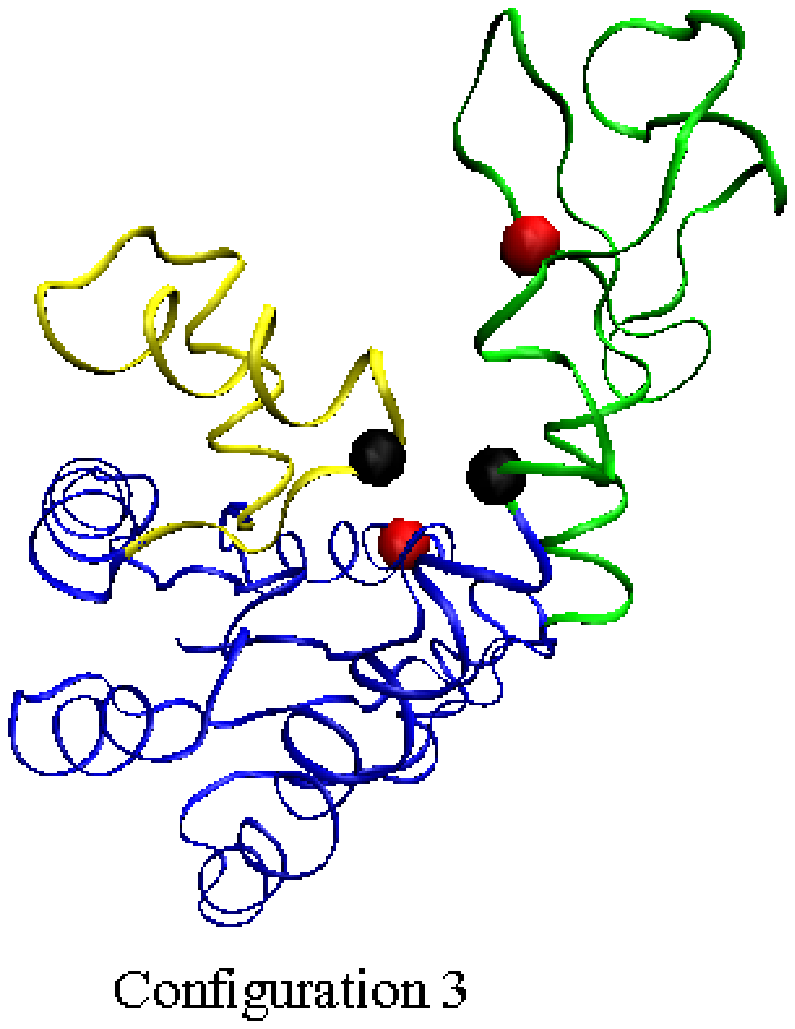}} &
\resizebox{2.0in}{!}{\includegraphics{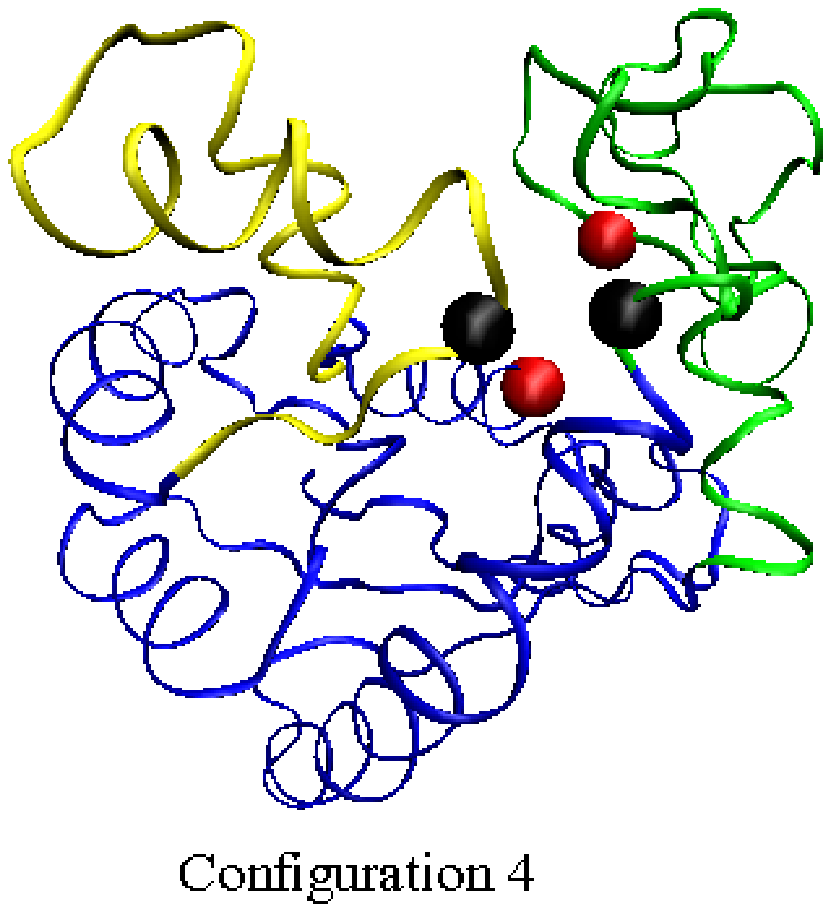}} \\
\end{tabular}
\caption{Four time--ordered, representative structures along the 
Open--BD--LID--Closed path obtained for Trajectory 3 in Figure~\ref{fig1} 
(Model 1). The CORE domain is blue, the BD is yellow, and the LID
is green. The configurations correspond to (296000, 802000, 822000, 1496000)
MC steps, respectively, in Trajectory 3 of Figure~\ref{DRMS_we_go}.}
\label{fig:1c}
\efig

\clearpage

\bfig
\begin{tabular}{cc}
\resizebox{2.0in}{!}{\includegraphics{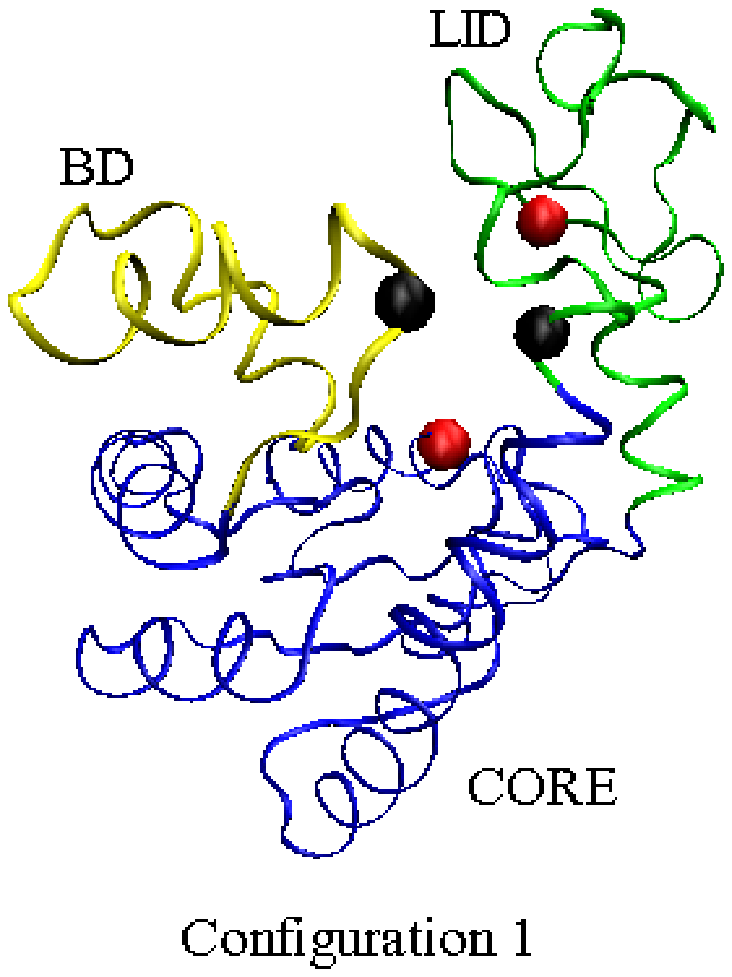}} &
\resizebox{2.0in}{!}{\includegraphics{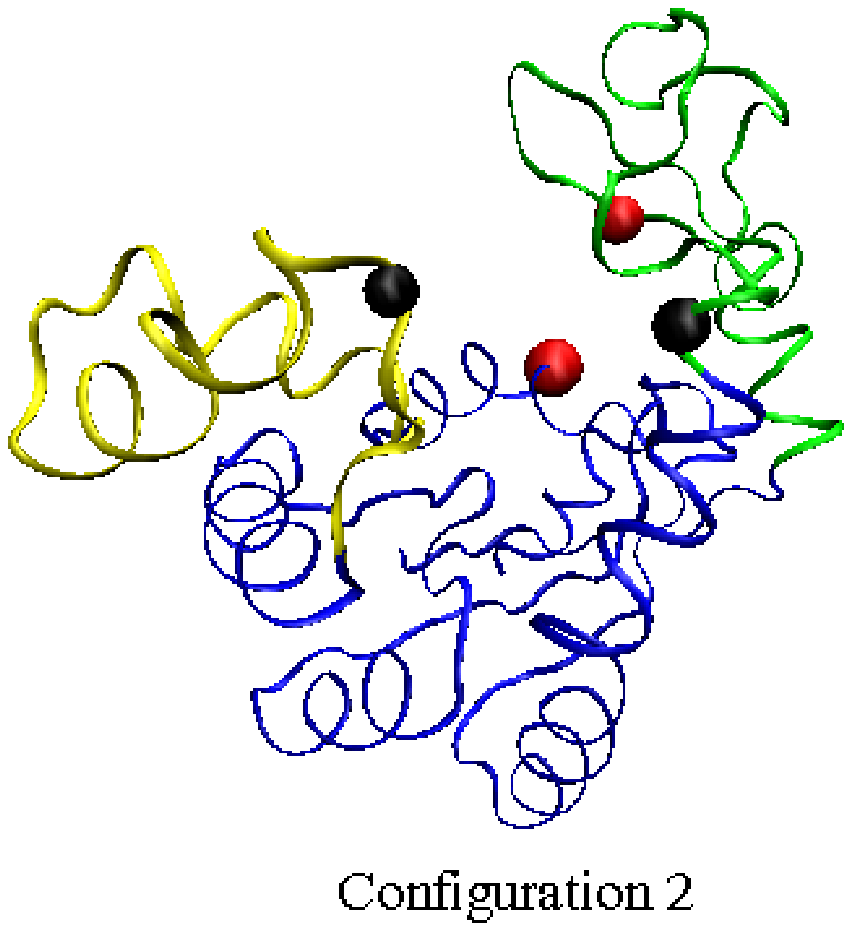}} \\
\resizebox{2.0in}{!}{\includegraphics{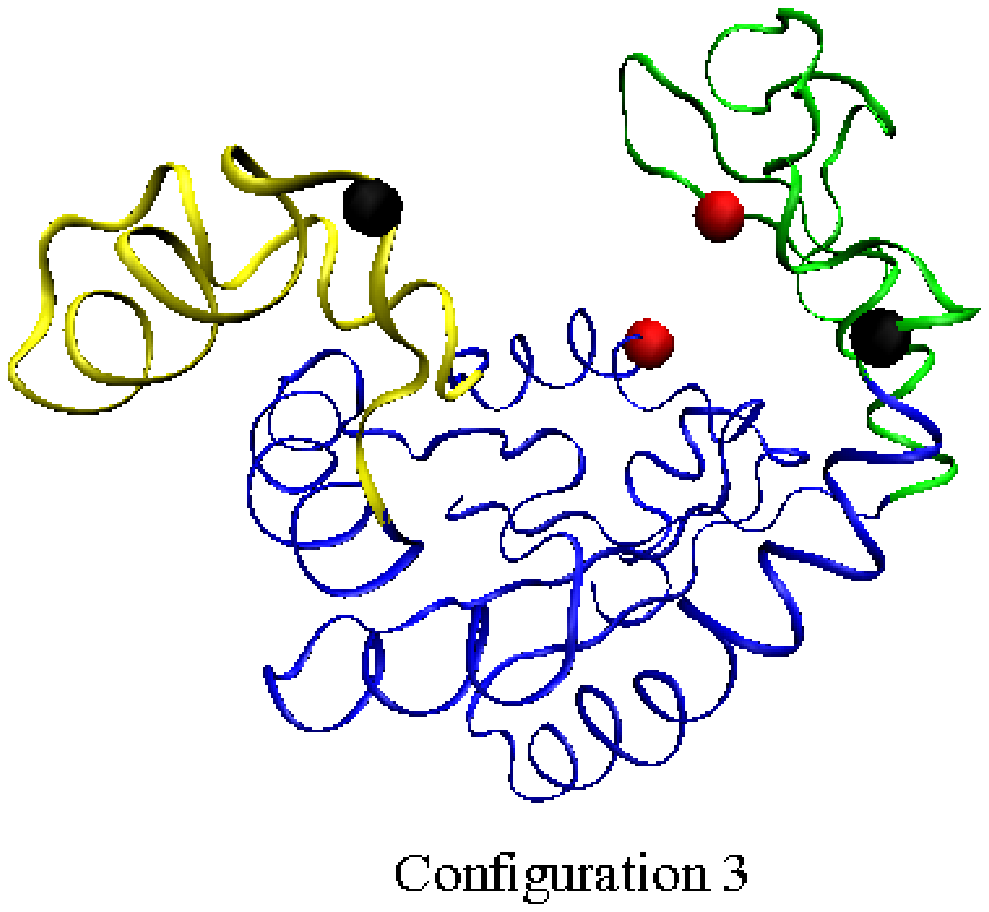}} &
\resizebox{2.0in}{!}{\includegraphics{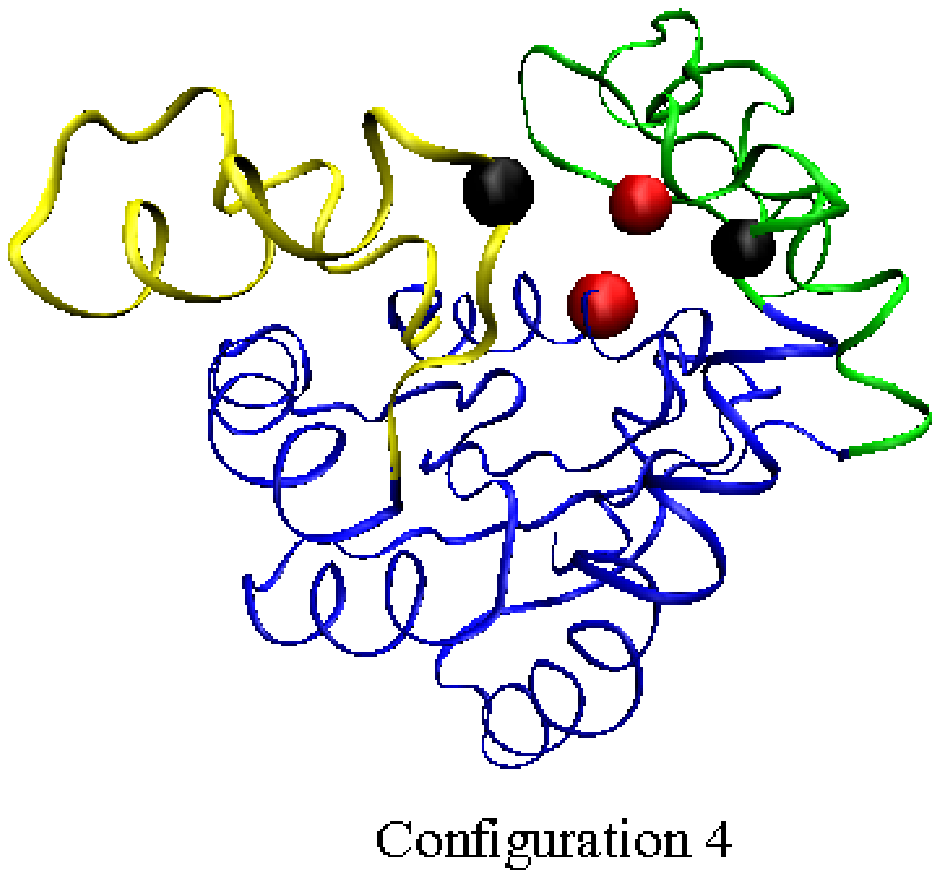}} \\
\end{tabular}
\caption{Four time--ordered, representative structures along the 
Open--LID--BD--Closed path obtained for Trajectory 1 in Figure~\ref{fig1} 
(Model 1). The CORE domain is blue, the BD is yellow, and the LID
is green. The configurations correspond to (128000, 206000, 306000, 480000)
MC steps, respectively, in Trajectory 1 of Figure~\ref{DRMS_we_go}.}
\label{fig:1d}
\efig

\clearpage

\bfig
\resizebox{3.5in}{!}{\includegraphics{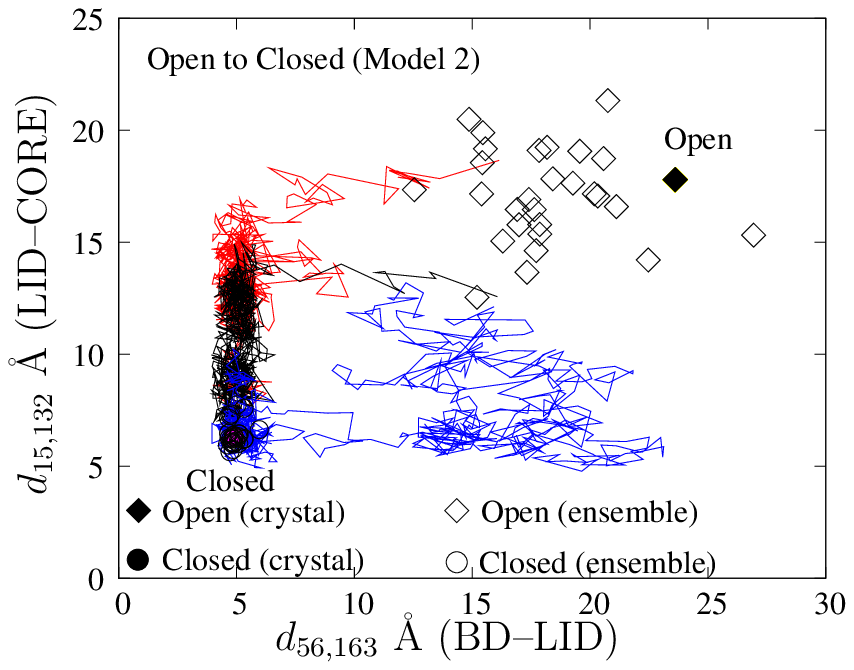}}
\caption{Path heterogeneity depicted via four typical trajectories
for the Open--to--Closed transition obtained
using Model 2 shown via distance between
the CORE and LID (ordinate) versus the distance between the BD and
the LID (abscissa). The ``dwell'' times for the trajectories in
the Open state excluded for clarity.}
\label{fig3}
\efig

\clearpage

\bfig
\resizebox{3.5in}{!}{\includegraphics{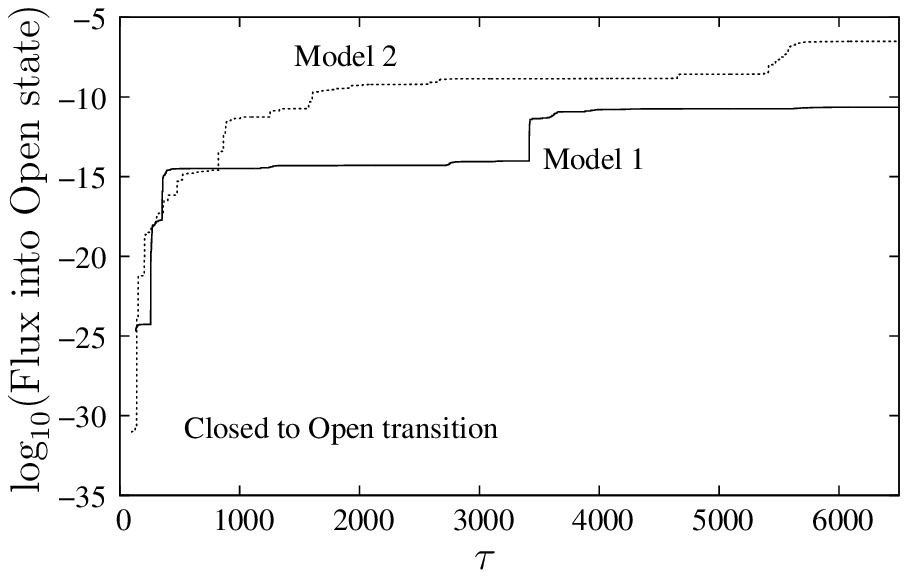}}
\caption{A comparison of probability fluxes into the Open state 
for two models as a function of
WE time increment (each $\tau$ is 2000 LBMC steps). The solid line is
for Model 1, whereas the dashed line is for Model 2. The log scale
emphasizes the small amount of flux in Closed--to--Open direction 
for both the models, as compared to fluxes in the reverse direction.}
\label{fig:rate_4ake_1ake}
\efig

\clearpage

\bfig
\resizebox{3.5in}{!}{\includegraphics{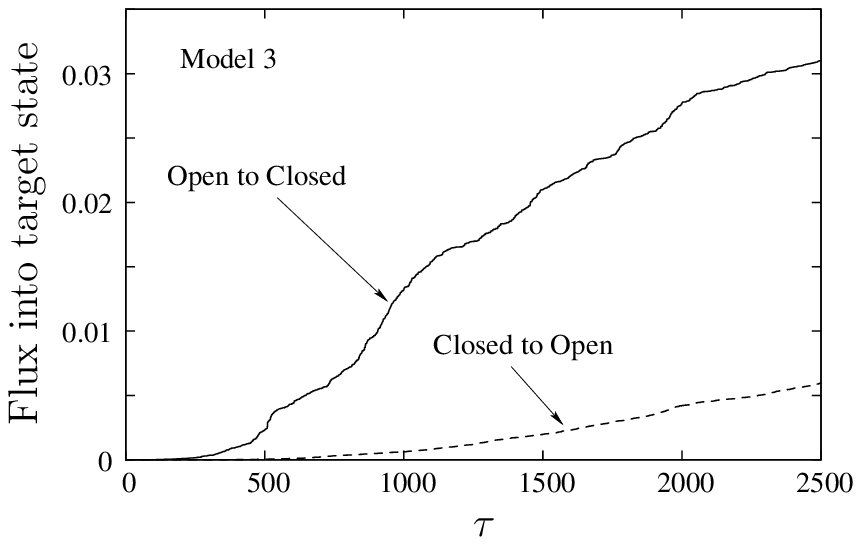}}
\caption{Probability flux in either direction for Model 3 as a
function of WE time increment (each $\tau$ is 2000 LBMC steps).
For this model, the Closed--to--Open flux is of the same order
of magnitude as in the reverse direction.}
\label{fig:rate_m3}
\efig

\clearpage

\bfig
\resizebox{3.5in}{!}{\includegraphics{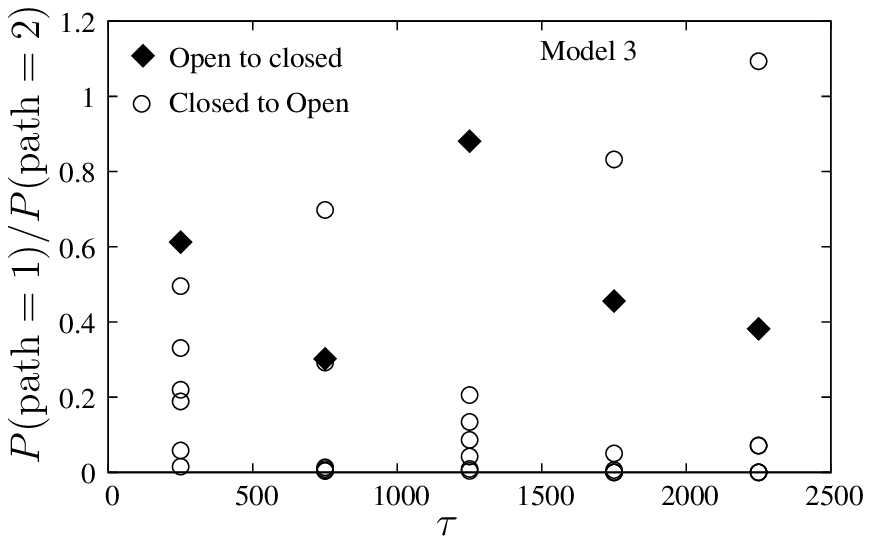}}
\caption{Ratio of probabilities of the two paths, 1 and 2, in the
Open--to--Closed (diamonds) and Closed--to--Open directions (circles).
Results from six independent WE simulations are shown in the
Closed--to--Open direction to highlight the fluctuations. All data
points are window averaged for 500 $\tau$.}
\label{fig:avgpath_m3}
\efig

\end{document}